\documentclass[aps,prl,10pt,twocolumn,showpacs,preprintnumbers,amsmath,amssymb]{revtex4-1}
\usepackage{comment}
\usepackage{graphicx,graphics,wrapfig,rotating}         
\usepackage{epsfig,subfigure}
\usepackage{dcolumn}                                    
\usepackage{bm,fancybox}
\usepackage{amsmath,amssymb}                          
\usepackage{times,euscript,oldgerm}              %
\usepackage[english]{babel}                      %
\usepackage{psfrag}
\usepackage{hyperref}
\usepackage{color}
\newcommand{\unit}[1]{\ensuremath{\,\mathrm{#1}}}
\newcommand{\Yb}{\ensuremath{^{171}\mathrm{Yb}^+}}

\newcommand{\MHz}{\unit{MHz}}

\newcommand{\THz}{\unit{THz}}

\newcommand{\micros}{\unit{\mu s}}
\newcommand{\ket}[1]{\ensuremath{\left|#1\right\rangle}}

\begin{document}

\title{Quantum Simulation and Phase Diagram of the Transverse Field Ising Model with Three Atomic Spins}
\author{E. E. Edwards$^1$, S. Korenblit$^1$, K. Kim$^1$, R. Islam$^1$,  M.-S. Chang$^1$, J. K. Freericks$^2$, G.-D. Lin$^3$, L.-M. Duan$^3$, and C.
Monroe$^1$}
\affiliation{$^1$ Joint Quantum Institute, University of Maryland Department of
Physics and \\
                    National Institute of Standards and Technology, College
Park, MD  20742 \\
             $^2$ Department of Physics, Georgetown University , Washington, DC
20057      \\
             $^3$ MCTP and Department of Physics, University of Michigan, Ann
Arbor, Michigan 48109}
\date{\today}
\begin{abstract}
We perform a quantum simulation of the Ising model with a transverse field using a collection of three trapped atomic ion spins.  By adiabatically manipulating the Hamiltonian, we directly probe the ground state for a wide range of fields and form of the Ising couplings, leading to a phase diagram of magnetic order in this microscopic system.  The technique is scalable to much larger numbers of trapped ion spins, where phase transitions approaching the thermodynamic limit can be studied in cases where theory becomes intractable.
\end{abstract}
\maketitle

At the pinnacle of quantum information science is the full scale quantum computer
\cite{nielsen}, where applications such as Shor's factoring algorithm \cite{Shor} can provide an exponential speedup compared with any known classical approach.  While large-scale quantum computers may not be available for some time \cite{Ladd}, more restricted quantum computers known as quantum simulators look promising right now \cite{Nori09}.  As first considered by Richard Feynman \cite{Feynman}, a quantum simulator controls interacting quantum bits (qubits) to implement evolution according to a known Hamiltonian \cite{Lloyd96}.  Then, by performing particular correlation measurements on the qubits, properties of certain Hamiltonians -- like their ground state -- can be extracted, often more efficiently than any classical simulation of the underlying quantum system \cite{Farhi01}.  A good example is a collection of interacting magnetic spins, where the Hamiltonian can easily be written down, yet the ground state of magnetic order cannot always be predicted, even with just a few dozen spins \cite{sandvik10}.

In this Letter we simulate the Ising model with an transverse magnetic field and generate a phase diagram, using a system of $N=$3 trapped atomic ions.  By adiabatically manipulating the Hamiltonian, we extract the phases of magnetic order in the ground state as a function of the transverse field and Ising couplings \cite{Schaetz08,Transverse,Kim10}.
While this system admits an exact theoretical treatment, it also represents the smallest possible system having multiple Ising couplings, which give rise to interesting magnetic order in the phase diagram.
Furthermore, the experiment is scalable to larger numbers of spins where theoretical predictions become intractable.

\begin{figure}[htp]
\begin{flushleft}
\subfigure{\includegraphics[width=1.0\linewidth]{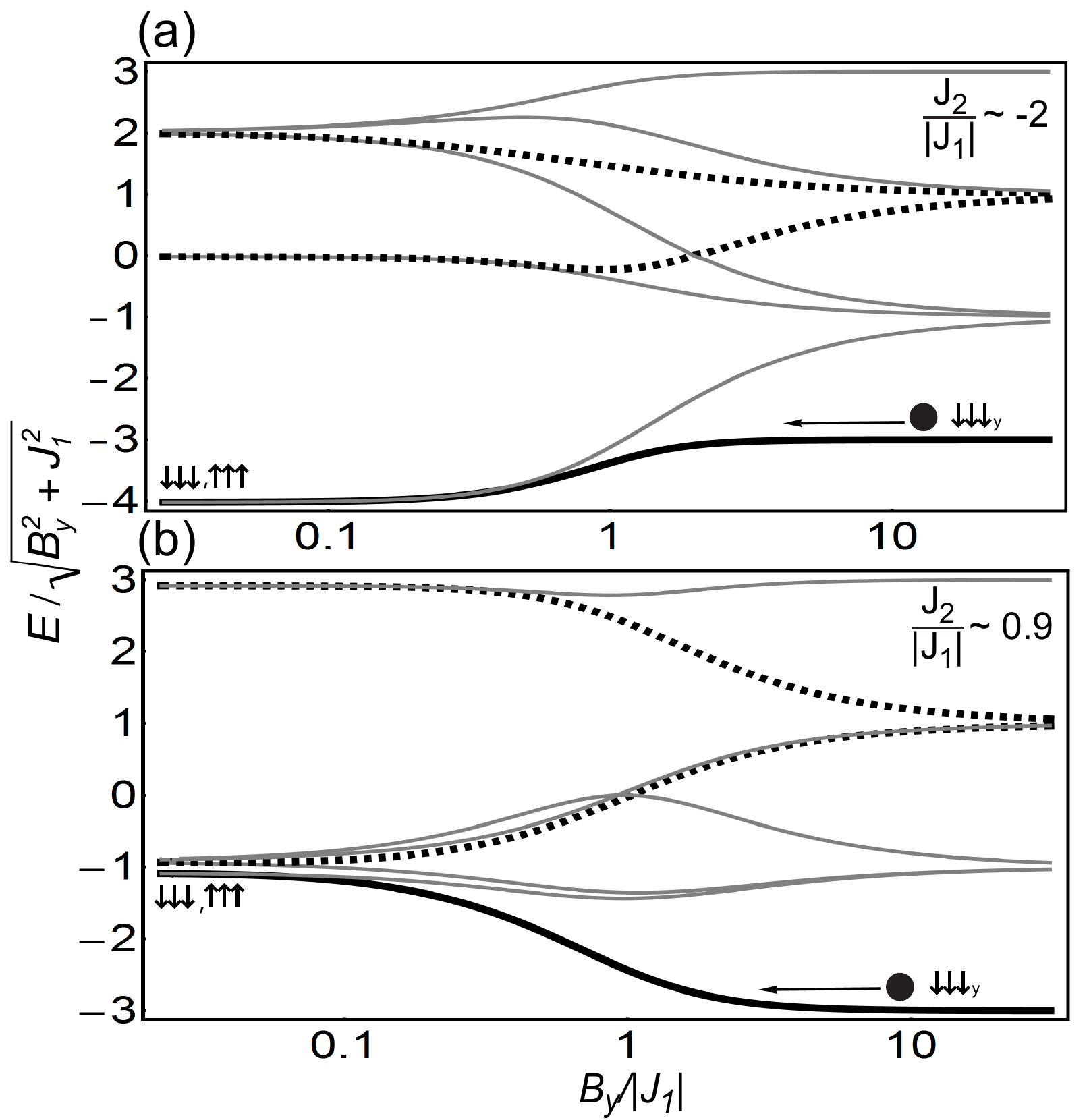}}
\end{flushleft}
\caption{Energy level diagrams for  Eq. \ref{Ham} with two different types of spin-spin interactions.  For both panels, the nearest-neighbor interactions are FM ($J_1<0$).  (a) The next-nearest-neighbor interaction is FM with $J_2/|J_1|\sim$ -2 and (b) AFM with $J_2/|J_1|\sim$0.9. The arrow in both diagrams indicates the trajectory in the simulation, initialized at $B_y/|J_1|\sim$10.  Under this condition, the initial ground state is an eigenstate of second term in Eq.  \ref{Ham} , a polarized state along $B_y$.  In both examples, at  $B\approx |J_1|$ some high energy states cross,  but  the ground state (black solid line) has no level crossings with any excited state.  Likewise, the highest energy state does not cross any other levels, allowing one to also adiabatically follow this state.  The dotted lines represent excited states which are coupled to the ground state along the path.  In the large field limit, the energy difference between ground and excited states $\Delta_{ge}$ (here, scaled by $\sqrt{B^{2}_y +J^{2}_1}$) is proportional to $B_y$, but as $B_y/|J_1|$  decreases the spin-spin couplings determine the energy difference and the form of the ground state. In both (a) and (b), the final ground state is FM (defined along the x-axis of the Bloch sphere), however in the case of (a), the gap to the nearest allowed excited state at the crossover is $\sim$15 times larger.}

\label{energy}
\end{figure}

The system is described by  the Hamiltonian 
\begin{equation}
H = \sum_{i < j} J_{i,j} \sigma_x^{(i)}\sigma_x^{(j)} + B_y\sum_i
\sigma_y^{(i)},
\label{Ham}  
\end{equation}
with Ising couplings $J_{i,j}$ between spins $i$ and $j$ and a uniform transverse magnetic field $B_y$. For three spins along a symmetric 1D chain,
we define $J_1 \equiv J_{1,2}=J_{2,3}$ as the nearest-neighbor interaction
strength and $J_2 \equiv J_{1,3}$ as the next-nearest-neighbor interaction, with
$\sigma_\alpha^{(i)}$ the Pauli spin operator of the $i$-th particle along the $\alpha$-direction.
Fig. \ref{energy} shows two energy spectra
as a function of the scaled transverse field $B_y/|J_1|$ in the case of ferromagnetic (FM) nearest-neighbor interactions ($J_1 < 0$).   We prepare the system in the ground state of the transverse field ($B_y \gg |J_1|$), depicted by
the solid circle (Fig. \ref{energy}), and then adiabatically lower the field compared to the Ising couplings. 
When $B_y/|J_1|\ll$1  the Ising interactions determine both the form of the ground state and the energy spacing  $\Delta_{ge}$ to the excited state(s). 

In Fig. \ref{energy}a, the next-nearest-neighbor interaction is also FM ($J_2<0$).
There are no level crossings with the ground state over the trajectory indicated by the arrow, thus if the evolution of the Hamiltonian is slow enough, the system remains in the ground state (solid black line).  We can also change the sign of $B_y$ and adiabatically follow the highest excited level, as it exhibits the same structure as the ground state.   In Fig. \ref{energy}b, $J_2>0$ and the next-nearest-neighbor interaction is antiferromagnetic (AFM).   The gap at the crossover to magnetic order defined by the Ising couplings is $\sim$15 times smaller than that of Fig. \ref{energy}a, requiring a slower change of $B_y/|J_1|$ to remain in the ground state.   

\begin{figure}
\subfigure{\includegraphics[width=1.0\linewidth]{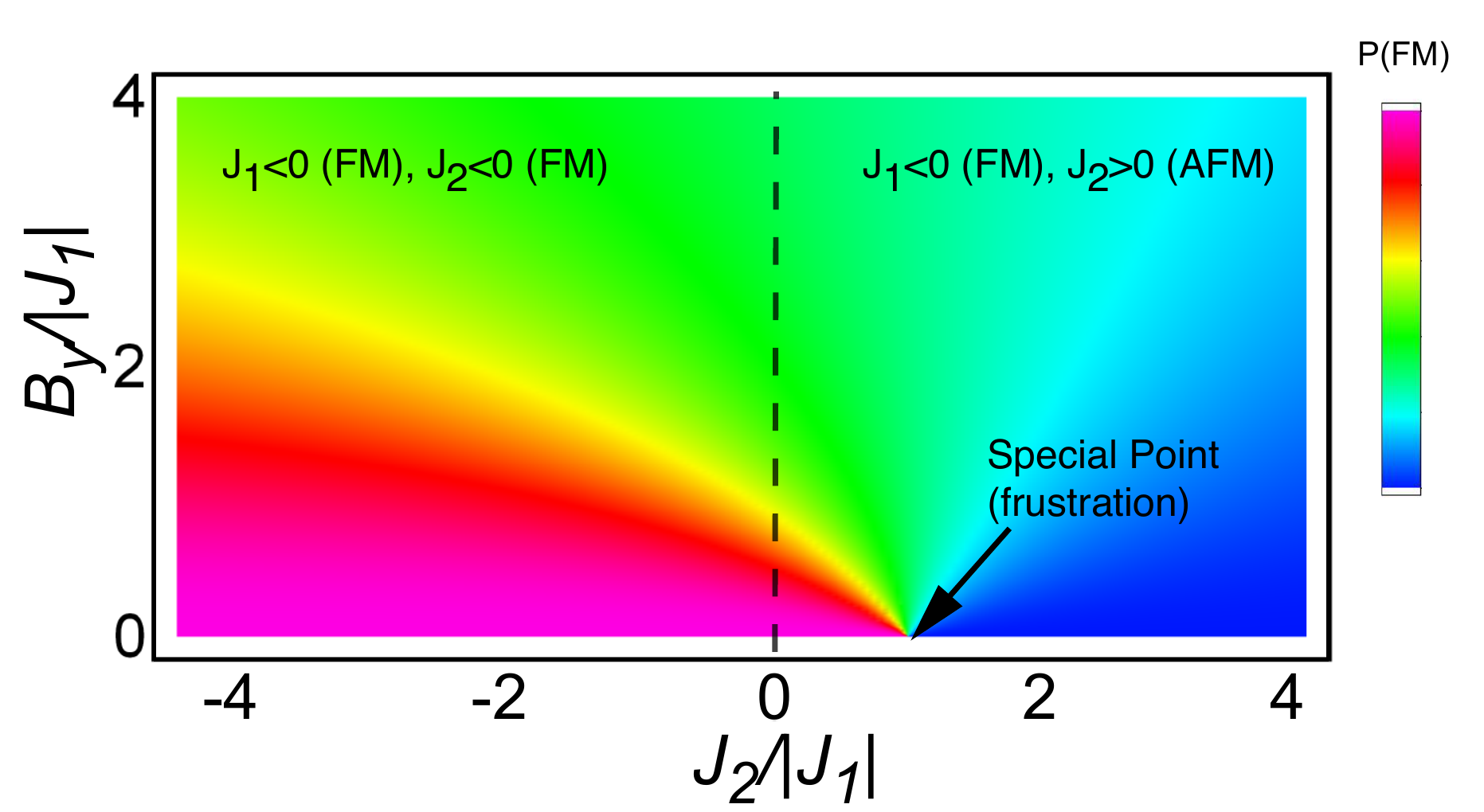}}
\caption{Theoretical phase diagram for Eq.  \ref{Ham}.  The color scale indicates the amplitude of  the FM order parameter, $P(FM)=P_{\ket{\uparrow\uparrow\uparrow}}+P_{\ket{\downarrow\downarrow\downarrow}}$.   Here,  $J_1$ is always negative, yielding FM order in that coupling. In the region where $J_2/|J_1|<$0, there is a crossover to FM order as $B_y/|J_1|$ is lowered (corresponds to Fig. \ref{energy}a).   When $J_2/|J_1|>$0, the  AFM and FM interactions compete (also shown in Fig. \ref{energy}b).  This gives rise to  a special sharpened point at $J_2/|J_1|=$-1 and $B_y=$0.  Here, the ground state is comprised of 6 states: four asymmetric  AFM and two FM states. } 
\label{PhaseAll}
\end{figure}

The competition between different parameters in Eq. \ref{Ham} gives rise to a complex phase diagram.  
The 2$^3$ possible spin configurations are defined as two FM states, $\ket{\uparrow\uparrow\uparrow}$ and $\ket{\downarrow\downarrow\downarrow}$, two symmetric AFM states,  \ket{\downarrow\uparrow\downarrow} and \ket{\uparrow\downarrow\uparrow}, and four asymmetric AFM states, \ket{\uparrow\uparrow\downarrow}, \ket{\uparrow\downarrow\downarrow}, \ket{\downarrow\uparrow\uparrow} and \ket{\downarrow\downarrow\uparrow}, all defined along the $x$-axis of the Bloch sphere.  
In Fig. \ref{PhaseAll}, we plot  a part of the theoretical
phase diagram where the nearest-neighbor interactions are always FM ($J_1<0$). The order parameter is the probability of occupying an FM state,  P(FM)~=~$P_{\ket{\uparrow\uparrow\uparrow}}+P_{\ket{\downarrow\downarrow\downarrow}}$.  
For regions where $B_y/|J_1|\gg$1, the ground state is polarized along $B_y$ with P(FM)$=1/2^{N-1}=1/4$.  As $B_{y}/|J_1|$ decreases, different magnetic phases arise.  
When the next-nearest-neighbor interaction is also FM ($J_2<0$), and $B_y/|J_1|\ll1$ the ground states are the two degenerate FM states (Fig. \ref{energy}a).  In the region where the next-nearest-neighbor interaction is AFM ($J_2/|J_1|>0$) and $J_2$ overpowers $J_1$,  the asymmetric AFM states are lowest in energy.   A special point appears at $J_2/|J_1|=-1$ and $B_y=0$, where all the contours of constant FM order meet.  Here, the ground state will be a superposition of the FM and
asymmetric AFM states. This effect arises because the pairwise interaction energy cannot be minimized individually, leading to a highly degenerate, or frustrated, ground state \cite{Kim10}.  

We experimentally simulate the transverse field Ising model of Eq. \ref{Ham} using cold trapped \Yb  ions \cite{porras04}, with the effective spin-$1/2$ system represented by the hyperfine ``clock"  states $^2S_{1/2}$ $\ket{F=1, m_F=0}$ and $\ket{F=0, m_F=0}$ 
denoted as $\ket{\uparrow}_{z}$ and $\ket{\downarrow}_{z}$, respectively \cite{Yb-qubit}, where $\ket{\uparrow}_z$=$\ket{\uparrow}+\ket{\downarrow}$ and  $\ket{\downarrow}_z$=$\ket{\uparrow}-\ket{\downarrow}$.  We confine $N=3$ atomic spins in a linear radiofrequency ion trap and couple them through $N$ collective transverse motional normal modes along one principal axis.  These vibrational normal modes, having frequencies $(\nu_1, \nu_2, \nu_3) \sim (4.334, 4.074, 3.674)$~MHz, are each cooled to near the ground state and deeply within the Lamb-Dicke limit \cite{Transverse, Kim10}. 

The effective magnetic field $B_{y}$, which produces Rabi oscillations between the two spin states, is generated by uniformly illuminating the ion chain with two Raman laser beams having a difference frequency at the hyperfine splitting, $\nu_{HF}=12.642821$ GHz.  For an individual beam detuning of ~$\sim$1.8$\THz$ below the $^2S_{1/2} - ^2P_{1/2}$ transition\cite{NIST} and a peak intensity of 10$\unit{W}/\unit{mm}^2$ each ion undergoes Rabi oscillations at a rate of 
$\Omega \sim 1\MHz$ and experiences a $\sim~20$~kHz differential AC Stark shift.  

The spin-spin interaction $J_{i,j}$ is created by coupling the ions' spin states through the normal modes of motion of the chain. The two Raman beams travel perpendicular to each other to have a wavevector difference $\delta k$ along the transverse direction.  The laser frequency of one of the two pathways is modulated to yield beatnotes (with respect to the non-modulated beam) at frequencies $\nu_{HF}\pm\mu$, imparting a spin-dependent force at frequency $\mu$ \cite{DidiRMP,MS}.  By controlling the beatnote detuning $\mu$, we tailor the Ising couplings according to
\cite{Transverse}

\begin{equation}
J_{i,j}=\Omega_i\Omega_j\sum_m{\frac{\eta_{i,m}\eta_{j,m}\nu_m}{\mu^2-\nu_m^2}}.
\label{coupling}
\end{equation}
Here,  $\Omega_i$ is the Rabi frequency of the i$^{th}$ ion.  The Lamb-Dicke parameter for the m$^{th}$ mode of the i$^{th}$ ion is  $\eta_{i,m}~=~b_{i,m}\delta k \sqrt{\hbar / (4\pi M \nu_m)}$,  where $b$  is the normal mode transformation matrix and $M$ the mass of a single ion.  In the above expression, we assume $\left|\nu_m-\mu\right|>>\eta\Omega_i$, so that phonons are only virtually excited. 

We initialize the spins along the $B_y$-direction through optical pumping 
($\sim$ 1$\micros$) and a $\pi/2$ rotation about the $-x$-axis of the Bloch sphere. 
The simulation begins with a simultaneous and sudden application of both $B_y$ and $J_{i,j}$ where $B_y$ overpowers $J_{i,j}$ ($B_y/|J_{i,j}| \approx 10$).  A typical experimental ramp of $B_{y}$ decays as $B_{y} = ae^{-t/\tau} + b$ with a time constant of $\tau~ \sim 30$ $\mu$s, varying from $a \sim 10$ kHz to a final offset of $b\sim500$ Hz after $t=300$ $\mu$s.  By varying the power in only one of the Raman beams, this procedure introduces a change in the differential AC Stark shift of less than 2 Hz.  We turn off the Ising interactions and transverse field
at different $B_y/|J_{i,j}|$ endpoints along the ramp.
We then measure the magnetic order along the $x$-axis of the Bloch sphere by first
rotating the spins by $\pi/2$ about the $y$-axis, and detecting the $z$-component of the spins through spin-dependent fluorescence \cite{Yb-qubit}. 
By repeating identical experiments $\sim 1000$ times, we obtain the probability for the system to be in a particular spin 
configuration.  We collect fluorescence with a photomultiplier tube, exhibiting $\sim97\%$ detection effeciency per spin after 0.8 ms exposure.  

\begin{figure}
\subfigure{\includegraphics[width=0.99\linewidth]{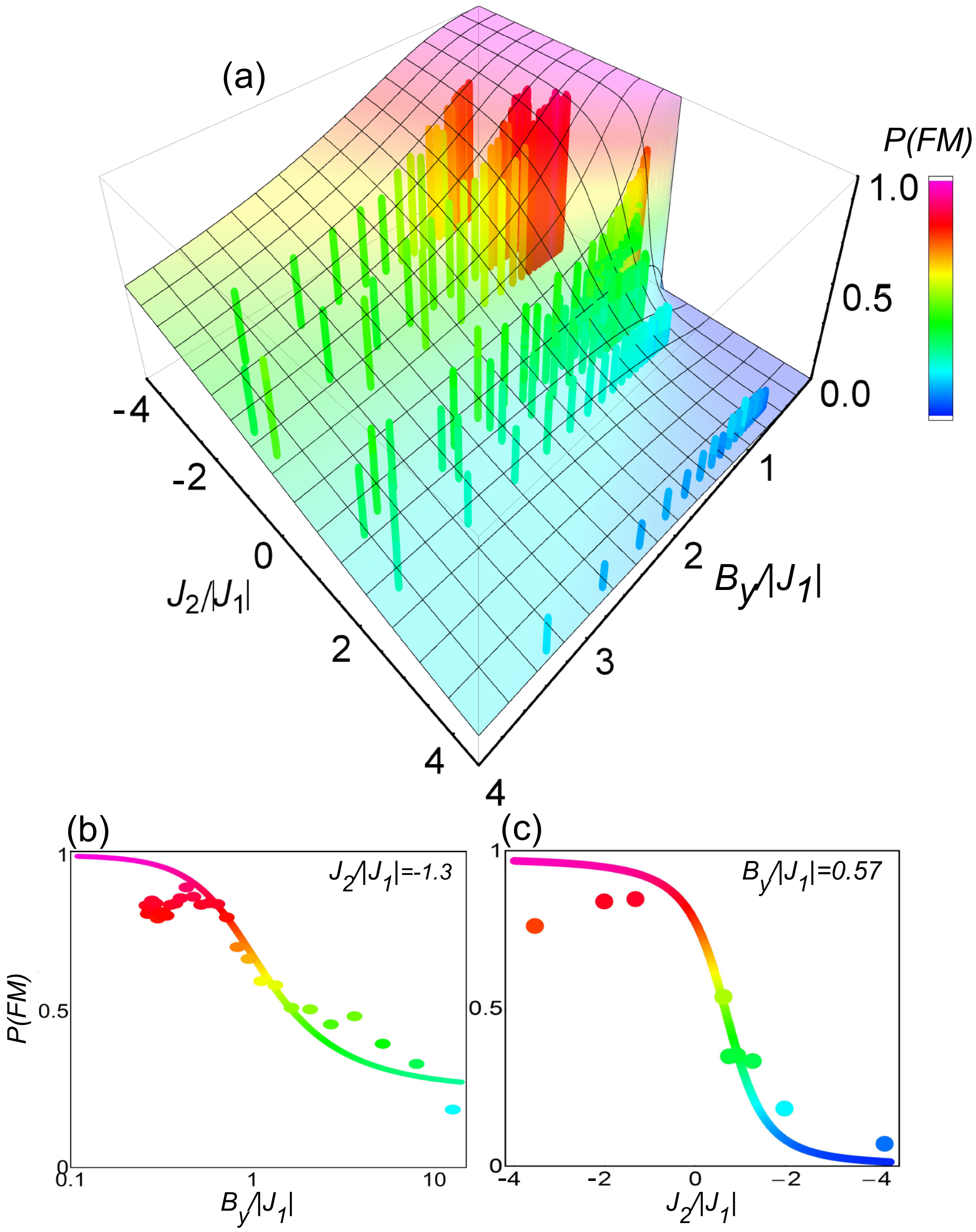}}
\caption{
Experimental measurements of the phase diagram for Eq.~ \ref{Ham} (solid bars) compared to the theoretical prediction from Fig. \ref{PhaseAll} (surface) .  The vertical amplitude is the FM order parameter P(FM)= $P_{\ket{\uparrow\uparrow\uparrow}}+P_{\ket{\downarrow\downarrow\downarrow}}$. The ratio of $B_y/|J_1|$ was varied from $\sim$10 to $\sim$0.1 for $J_2/|J_1|$ values of -1.3,-2.0, -3.6, 4.2, 2.0, 1.3, 0.92, 0.74, and 0.62.  $J_1<$ 0 for all traces.  (b) As $B_y/|J_1|\rightarrow$~0 in the region where $J_2/|J_1|<-1$ , we observe a smooth crossover to FM order.  The filled circles and solid line are the data and theory for $J_2/|J_1|=-1.3 $, respectively (c) When changing $J_2$ for a fixed and small value of $B_y/|J_1|$ the system undergoes a sharp transition.  The data (filled circles) shown is for a scan of $B_y/J_y=0.57$.  The average deviation per scan of $B_y/|J_1|$ from the exact ground state is~ $\sim0.09$.}
\label{specialpoint}
\end{figure}

This procedure is performed for nine different combinations of $J_1$ and $J_2$ set by the beatnote detuning $\mu$ from Eq.~ \ref{coupling}. In  Fig. \ref{specialpoint} we present the results as a 3D plot of the FM order parameter, with the theoretical phase diagram (surface) in Fig. \ref{PhaseAll} superimposed on the data.   The data is in good agreement with the theory (average deviation per trace is $\sim0.09$) and shows many of the essential features of the phase diagram.  As $B_y/|J_1|$ decreases, a smooth crossover from a non-ordered state to FM order occurs in the region where $J_2/|J_1|<0$ (Fig. \ref{specialpoint}b).  As the number of spins increases, this is an example of a quantum phase transition.  A first order transition due to an energy level crossing is apparent (Fig. \ref{specialpoint}c) when changing $J_2$ for a fixed and small value of $B_y/|J_1|=0.57$.  This transition is sharp, even in the case of three spins.    
The data (e.g. Fig. \ref{specialpoint}b) show small amplitude oscillations in the initial evolution due to the sudden application of the spin-spin interaction, which is held constant during the simulation to minimize variation in the differential AC stark shifts.  This limitation can be removed by choosing the Raman laser detuning such that contributions from the $^2P_{3/2}$ energy level lead to a minimum in the ratio of the differential AC stark shift to the resonant Rabi frequency \cite{Wes10}.  


We now investigate adiabaticity of the Hamiltonian trajectory $H(t)$, 
characterized by the condition \cite{Schiff}
\begin{equation}
\frac{\dot{B_y}(t)\epsilon}{\Delta_{ge}^2}\ll1
\label{adiabatic}
\end{equation}
In this expression, the dimensionless quantity
$\epsilon$~$\equiv$~$\left\vert \langle g(t)|\frac{dH(t)}{dB_y}|e(t)\rangle\right\vert$
characterizes coupling from the ground state $\ket{g(t)}$ to any excited state $\ket{e(t)}$ with energy gap $h \Delta_{ge}$.  This parameter is small, of order unity for this simulation, but is peaked at a crossover in magnetic order, where the instantaneous eigenstates are most rapidly varying.  Therefore, Eq. \ref{adiabatic} states that to remain adiabatic, the slope of the time-dependent $B_y$-field profile must be shallow when the gaps in the energy spectrum are small (as in Fig. \ref{energy}b), in particular near a crossover (phase transition for large $N$).  

\begin{figure}
\includegraphics[width=1.0\linewidth]{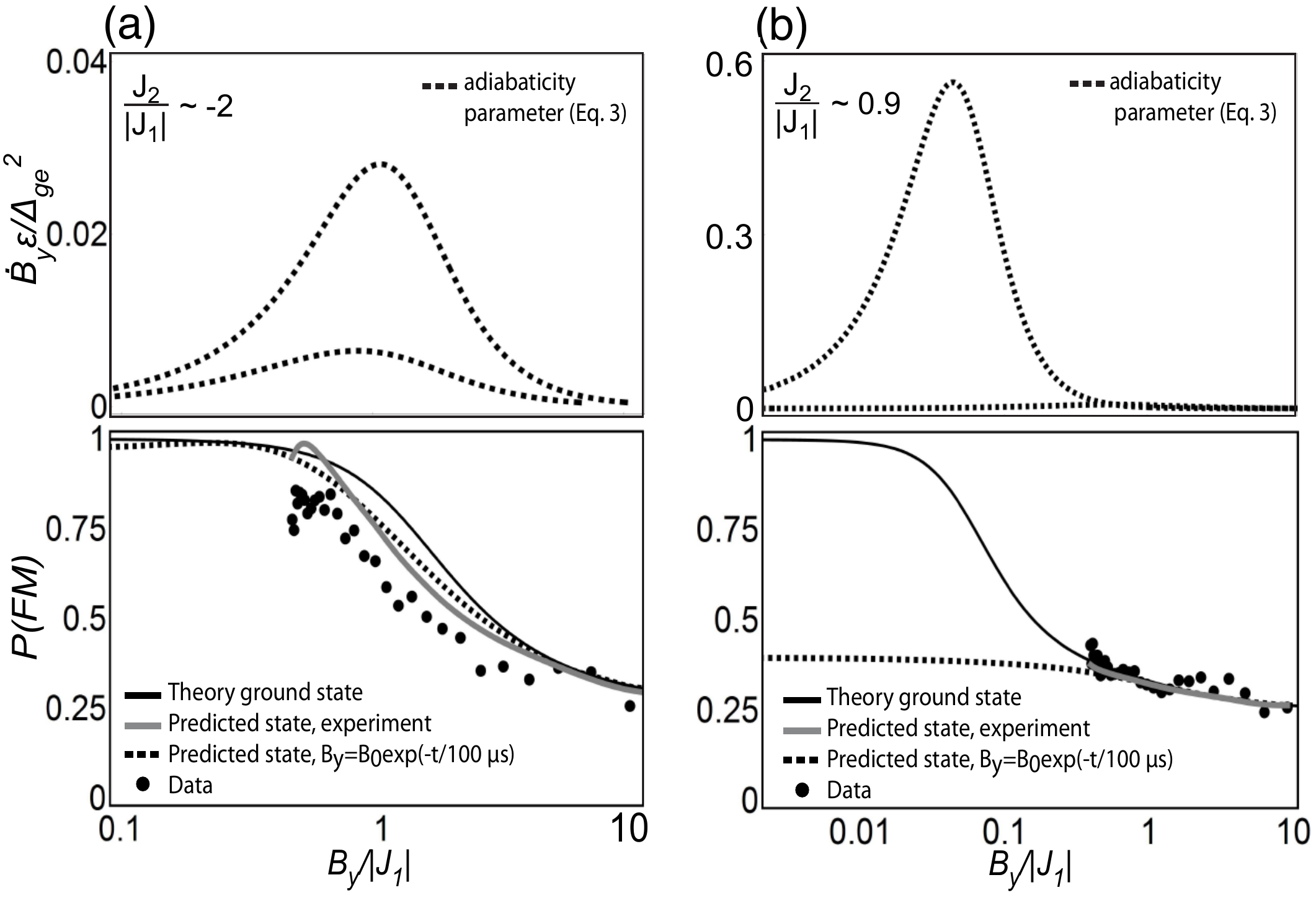}
\caption{
Adiabaticity for two cuts in the phase diagram where (a) $J_2/|J_1|=-2$; and (b) $J_2/|J_1|=0.9$.  The upper plots show the theoretical ground state adiabaticity parameter $\dot{B_y}(t)\epsilon/\Delta_{ge}^2$ of Eq.~ \ref{adiabatic} (dashed lines) for each of the two coupled excited states (see Fig. \ref{energy}).  We use a pure exponential decay ramp for the field $B_y$ with time constant $100$ $\micros$ in order to match the experiment for large values of $B_y/J_1$ while also extending the theory curves below the minimum value of $B_y$ reached in the experiment.   At every point in the the trajectory of (a), we find that the adiabatic condition (Eq. \ref{adiabatic}) is satisfied for typical experimental times (300 $\micros$). On the other hand, in (b), there is a significant probability of diabatic transitions to excited states for $B_y/J_1\ll1$. In the lower plots, we compare the observed FM order parameter (points) with theory.  In (a), the theoretical order from the exact experimental ramp with a $35$ $\micros$ time constant and final offset value given in the text (grey solid line) is in reasonable agreement with the order in the true ground state (black solid line) for $B_y/J_1>0.5$. The dotted line is the expected state evolution for a pure exponential decay ramp with a $100$ $\micros$ time constant, allowing $B_y\rightarrow0$. In (b), the data also matches well to theory, as we avoided the regions where diabatic transitions are expected for $B_y/J_1\ll1$.  According the calculations,  the duration of three-spin experiments near the special point should be on the order of milliseconds.  
}
\label{evolution}
\end{figure}

In Fig. \ref{evolution}, we investigate this adiabatic criteria for two different types of next-nearest-neighbor coupling.   In Fig. \ref{evolution}a all interactions are FM and  $J_2/|J_1|\sim$ -2 (as in Fig. \ref{energy}a).  The dashed lines in the top panel are the adiabaticity parameter from Eq. \ref{adiabatic} calculated over the trajectory for the two coupled excited states (recall Fig. \ref{energy}).  Due to the 500~Hz final offset of $B_y$, the simulation stops at  $B_y/|J_1|\sim0.5$.  To examine the behavior extended below this value, we calculate the criteria for an exponential ramp with a 100~$\micros$ time constant.  This profile was chosen to overlap with experimental parameters for large $B_y/|J_1|$ and also reach $B_y=0$ in a typical simulation time ($\sim 300 \micros$).   The results indicate that Eq. \ref{adiabatic} is satisfied over the trajectory;  $\dot{B_y}(t)\epsilon/\Delta_{ge}^2$ remains much less than one even with a maximum occurring at $B_y/|J_1|\sim 1$.  To demonstrate the simulation is indeed adiabatic for these parameters, the measured probability P(FM) (solid dots) is shown in the lower panel of  Fig.~ \ref{evolution}a.  The black line represents the adiabatic ground state and the grey line is the theoretical expected probability including the experimental ramp.   The dotted line in this figure is the theoretical state evolution using a $B_y$-field ramp that reaches zero.   The predicted evolution does not significantly deviate from the ideal ground state and the data is in good agreement with all three theory curves.  

Fig. \ref{evolution}b presents the case when the next-nearest-neighbor interaction is AFM and $J_2/|J_1|\sim~ $0.9 (as in Fig.\ref{energy}b).   When $B_y/|J_1|\ll1$,  $\dot{B_y}(t)\epsilon/\Delta_{ge}^2$ reaches a maximum value of $\sim0.6$,  indicating that the probability for excitations will likely increase.  This difference is because in this case the gap $\Delta_{ge}$ at the 'critical' point is $\sim$~15 times smaller than that in Fig. \ref{energy}a.   In contrast to the FM $J_2$ case, the theoretical probability curves shown in the lower panel of Fig. \ref{evolution}b predict significant diabatic effects when using this $B_y$-field profile for simulations near the special point.  In fact, to successfully evolve to the true ground state near $B_y=0$, the simulation time (assuming same initial conditions and an exponential ramp of $B_y$) should be at least a factor of ten longer.  

Because all the data lies outside of the region where the energy gaps are small, the diabatic excitations are minimal, but further experimental study is needed to precisely quantify this effect. One method to probe excitations, which may also be useful as $N\gg1$, is to perform and then reverse the experimental ramp and measure the probability of returning to the initial state.  The main contributions to the overall data offset from theory shown in Fig. \ref{evolution} and Fig. \ref{specialpoint} are spontaneous emission due to off-resonant scattering  (probability $\sim 15\%$ in 1 ms), imperfect optical pumping (state preparation), parasitic fields along the $x-$ and $z-$axes, and state detection error.  Additional phonon terms not appearing in the Hamiltonian of Eq. \ref{Ham} are expected to contribute at a level under 2$\%$ \cite{Transverse}.

As the number of spins $N$ grow, the technical demands on the apparatus are not forbidding \cite{Kim10,Transverse}.   In particular, the expected adiabatic simulation time for this model is inversely proportional to the 'critical' gap in the energy spectrum;  for the transverse field Ising model in a finite-size system, this gap decreases as $N^{-1/3}$ \cite{Caneva07}.  Scaling this system to accommodate long ion chains (approaching the thermodynamic limit) will allow investigation of behavior near critical points.  This is
interesting for $N>$20, where general spin models become theoretically intractable.  For instance, the Lanczos algorithm \cite{Lanczos} can be used to find low-lying states of  a 30 spin/site problem if the $2^{30}$ element matrix is sparse.  If one is interested in a non-sparse matrix, as is the case for the ion system with long-range
magnetic coupling, the limiting number of spins is $\sim$20.  Theoretical investigations of dynamics limit the number of spins further \cite{sandvik10,30sites}.   We note that this approach to quantum simulation is versatile and may be extended to simulate Heisenberg or XYZ spin models using additional laser beams \cite{porras04}.

\setcounter{secnumdepth}{0}

\begin{acknowledgments}
This work is supported under Army Research Office (ARO) Award W911NF0710576 with
funds from the DARPA Optical Lattice Emulator (OLE) Program, IARPA under ARO
contract, the NSF Physics at the Information Frontier Program, and the NSF
Physics Frontier Center at JQI.

\end{acknowledgments}
\bibliography{phase_prl}

\end{document}